\documentclass[aps,preprint,showpacs,amsmath,amssymb]{revtex4}
\usepackage{amsmath}
\usepackage{txfonts}
\usepackage{mathrsfs}
\usepackage{graphicx}
\usepackage{dcolumn}
\usepackage{bm}
\begin{document}
\title{Extension of Renormalizability}
\author{S.S.Wu}
\email{wuss@jlu.edu.cn}
\author{H.B.Yao}%
\author{Y.J.Yao}
\affiliation{%
Center for Theoretical Physics and School of Physics, JiLin
University, Changchun 130023, P.R.China
}%
\date{\today}
\begin{abstract}
 Arguments are provided which show that extension of
renormalizability in quantum field theory is possible. A dressed
scheme for the perturbation expansion is proposed. It is proven that
in this scheme a nonrenormalizable interaction becomes
renormalizable in the restrictive sense, i.e. its ultraviolet
divergences can be cancelled by a finite number of counterterms
included in the Lagrangian. As an illustration, the renormalization
of the $\pi$-nucleon pseudovector interaction is discussed in some
detail.
\end{abstract}
\maketitle
 It is now generally believed that renormalizability is
not a fundamental requirement of quantum field theory. In fact, the
widely acknowledged effective field theory \cite{w1, w2, georgi}
contains nonrenormalizable interactions. It has been especially
emphasized by Weinberg \cite{w1} that renormalizability is
unnecessary for the following main reasons: (1) it places a too
stringent restriction on the possible types of renormalizable
interactions and (2) as regards the cancellation of ultraviolet (UV)
divergences, nonrenormalizable theories are actually also
renormalizable, if all of the possible interactions allowed by
symmetries are included in the Lagrangian, because then there will
be enough counterterms to cancel every UV divergence. However, it is
still desirable to find means to broaden the extent of
renormalizability, since for a renormalizable interaction only a
finite number of counterterms in the Lagrangian is needed for the
elimination of infinities, while an infinite number is necessary, if
it is a nonrenormalizable (NR or nr) interaction (I). Hereafter we
shall always understand renormalizability in the above restrictive
sense specified by finite number. We would like to show that such an
extension is indeed possible. In this letter we shall only consider
ordinary quantum field theory based on special relativity. Consider,
for instance, a fermion propagator
\begin{align}
G_{\alpha\beta}(x=x_1-x_2)=<T[\psi_\alpha(x_1)\overline{\psi}_\beta(x_2)]>
=\int\frac{d^{4}k}{(2\pi)^4}e^{ikx}G_{\alpha\beta}(k), \label{eq1}
\end{align}
where $kx=k_{\mu}x_{\mu}=\vec{k}\cdot\vec{x}-k_0t$  and $
k_\mu\equiv(\vec{k},ik_0)$. Let the superscript "0" indicate a
zeroth order approximation. The Dyson-Schwinger equation for
$G_{\alpha\beta}(k)$ reads
\begin{subequations}
\renewcommand{\theequation}{\theparentequation-\arabic{equation}}
\begin{equation}
G(k)=G^0(k)+G^0(k)\Sigma(k)G(k).\label{eq2.1}
\end{equation}
If an appropriate approximation $\Sigma_d(k)$ to the self-energy
$\Sigma(k)=\Sigma_d(k)+\Sigma_r(k)$ has been found, we may introduce
a dressed propagator $G_d(k)$ and rewrite Eq.(\ref{eq2.1}) as
follows
\begin{align}
G_d(k)&=G^0(k)+G^0(k)\Sigma_d(k)G_d(k),\label{eq2.2}\\
G(k)&=G_d(k)+G_d(k)\Sigma_r(k)G(k).\label{eq2.3}
\end{align}
\end{subequations}
According to perturbation theory it is not difficult to see that a
perturbation series can also be expanded in terms of $G_d(k)$
(dressed scheme, DS) instead of $G^0(k)$ (ordinary scheme, OS), if
proper care has been taken to avoid redundancy of diagrams. Clearly,
the same remark also applies to boson propagators. Consider an
arbitrary connected one-particle irreducible Feynman diagram F. Let
us assume that each interaction $i $ in the Lagrangian is
characterized by $n_{i\kappa}$ fields of type $\kappa$ and $d_i$
derivatives acting on these fields. Following the argument given in
\cite{w1,muta}, one finds easily that the superficial degree of
divergence $d_F$ of diagram F can be written in the form:
\nocite{bracco}\nocite{bielajew1} \nocite{bielajew2}
\begin{subequations}
\renewcommand{\theequation}{\theparentequation-\arabic{equation}}
\begin{align}
d_F&=4-\Sigma_\kappa E_\kappa (2-p_\kappa)-\Sigma_iN_ir_i-\Sigma'_id_i, \label{eq3.1}\\
\tilde{d}_F&=d_F+\Sigma'_id_i \label{eq3.2},\\
r_i&=4-d_i-\Sigma_\kappa n_{i\kappa}(2-p_\kappa)\label{eq3.3},
\end{align}
\end{subequations}
 where we have expressed the asymptotic behavior of the propagator
$\Delta_\kappa(k)$ of field $\kappa$ (except the fermion propagator
which is denoted by G ) $\Delta_\kappa(k)\sim k^{-2p_\kappa}$,
$E_\kappa $ is the number of external lines of field $\kappa $,
$N_i$ the number of vertices of interaction $i$ in F and $r_i$ is
defined by Eq.(\ref{eq3.1}), where the prime over $\Sigma_i$ means
that $i$ only runs over those vertices
 which are connected with external lines and whose momentum factor
 becomes an external momentum (see Eq.(\ref{eq8})).  Since external momenta are
 not involved in the momentum integration, their
 $d_i$-contribution to $d_F$ should be subtracted. However, if we consider the
 asymptotic behavior of diagram $F$, it is given by $\tilde{d}_F$ (see
 Eq.(\ref{eq3.2}))\cite{w1}, because all the external momentum factors should
 be included. Consider, for instance, the pseudovector
$\pi$-N interaction (PVI)
$\mathscr{L}_{pv}=if\overline{\psi}\gamma_\mu\gamma_5\vec{\tau}\cdot(\partial_\mu\vec\phi)\psi$,
 we have $d_{pv}$=1, and $G^0(k)\sim k^{-1}$, $\Delta^0_\pi(k)\sim k^{-2}$, or $p_{_N}=1/2$ and $p_\pi=1$
 if we assume the ordinary scheme (OS). According to Eq.(\ref{eq3.3})
 $r_{pv}=-1$. Eq.(\ref{eq3.1}) says that $d_F$ grows with $N_i(i=pv)$, thus
 as is wellknown, PVI is nonrenormalizable. Now let us study DS. The one loop
 approximation to the nucleon self-energy reads
\begin{figure}
\fbox{\includegraphics{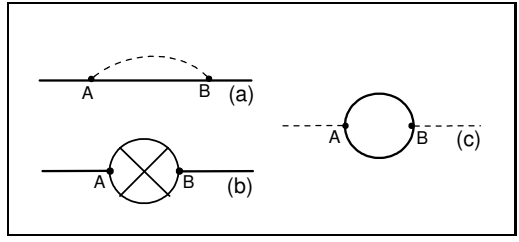}}
\caption{\label{fig1}Fermion self-energy (a) and (b) contributed by
PVI and CTL, respectively. (c) $\pi$-meson self-energy.}
\end{figure}
\begin{equation}
\Sigma_{pv}(k)=3f^2\int\frac{d^4q}{(2\pi)^4}\gamma_\mu q_\mu\gamma_5
G(k-q)\gamma_\nu q_\nu\gamma_5 \Delta_\pi(q).\label{eq4}
\end{equation}
It is known that we have $\Sigma(k)=\gamma_\mu k_\mu a(k^2)-i M
b(k^2)$. Substituting
\begin{equation}
G^0(k)=-[\gamma_\mu k_\mu-i(M-i\varepsilon)]^{-1},
\Delta^0_\pi(k)=-i[k^2+m_\pi^2-i\varepsilon]^{-1}\label{eq5}
\end{equation}
in Eq.(\ref{eq4}), one finds that $\Sigma_{pv}(k)$ is divergent.
However, as usual, one may calculate it by means of regularization,
namely by Feynman's parametrization and dimensional regularization
or by the counterterm method, i.e. by introducing counterterms in
the Lagrangian and eliminating the divergences by cancellation. From
both of these methods one obtains that the regularized $a_{pv}(k^2)$
and $b_{pv}(k^2)$ behave asymptotically as
\begin{equation}
a_{pv}(k^2)\sim k^2, \    b_{pv}(k^2)\sim k^2 \label{eq6}
\end{equation}
(see Eqs.(10) and (\ref{eq16}-17)). A formal solution to Eq.(2) can
be written in the form
\begin{align}
G(k)=-[\gamma_\mu k_\mu-i M+\Sigma(k)]^{-1}. \label{eq7}
\end{align}
Let us choose $\Sigma_d(k)=\Sigma_{pv}(k)$. From Eq.(\ref{eq2.2}) we
obtain $G_d(k)$. According to Eqs.(\ref{eq6}) and (\ref{eq7}) we
assert $G_d(k) \sim k^{-3}$ which implies $p_{_N}=3/2$. Thus, if the
expansion is in terms of $G_d(k)$, from Eq.(\ref{eq3.2}) we get
$r_{pv}=1$, even though $\Delta_\pi (k)$ is still taken as
$\Delta^0_\pi (k)$. Eq.(\ref{eq3.1}) shows that now PVI becomes
super renormalizable. Since $G_d(k)$ is derived from $G^0(k)$
through consideration of higher order terms (in fact, through the
Dyson-Schwinger equation, Eq.(\ref{eq2.2}), it has summed an
infinite series produced by $G^0$ and $\Sigma_d$ ), one may feel
strange why there is such a drastic change of convergence behavior
between OS and DS. This is clearly due to the fact that
$\Sigma_d=\Sigma_{pv}$ is divergent, PVI is nonrenormalizable (see
below) and we have made $\Sigma_{pv}$ finite by regularization.
Hence DS may differ from OS essentially. However, for example, for
the renormalizable pseudoscalar $\pi$-N interaction the one loop
contribution to the nucleon self-energy yields $a_{ps}(k^2) \sim \ln
k^2$ and $b_{ps}(k^2)\sim \ln k^2$, therefore $G_d(k)\sim (k\ln
k^2)^{-1}$ as given by Weinberg's theorem. This shows that $p_{_N}$
is essentially equal to $1/2$ and no unexpected change of
renormalizability occurs. Let us now consider the $\pi-$meson
self-energy. The one loop contribution is
\begin{equation}
\Pi(k)=-3f^2\int\frac{d^4q}{(2\pi)^4}Tr[\gamma_\mu k_\mu\gamma_5
G(k+q)\gamma_\nu k_\nu\gamma_5 G(q)].\label{eq8}
\end{equation}
It is seen that the momentum factors $k$ coming from derivatives
acting on the $\pi-$fields now belong to external momenta, thus they
do not contribute to the superficial degree of divergence $d_\pi$ of
$\Pi(k)$ as mentioned in Eq.(\ref{eq3.1}). In OS $d_\pi=2$ and
$\Pi_\pi(k)$ is divergent. Hence, for its calculation regularization
is necessary. However, in DS since in Eq.(\ref{eq8}) G should be
replaced by $G_d$, whose asymptotic behavior is $G_d(k)\sim k^{-3}$,
we have $d_{d\pi}=-2$, thus $\Pi_d(k)$ is convergent.

From Eq.(\ref{eq2.2}) it is seen that Eq.(\ref{eq4}) is an integral
equation for $\sum_{pv}(k)$. Whether its solution exists concerns
the self-consistency of our method. In the case of renormalizable
interactions such integral equation has been considered previously
\cite{brown, bracco, bielajew1, bielajew2, wuss}. Here,we shall only
consider the case where the asymptotic behavior of $G(k)\sim
k^{-2p_{_N}}$ with $1\le2p_{_N}\le4$. The right-hand side of
Eq.(\ref{eq4}) is then divergent and has to be regularized. Our
previous result indicated in Eq.(\ref{eq6}) corresponds to a
solution to Eq.(\ref{eq4}) obtained by a first iteration with the
initial input given by Eq.(\ref{eq5}). Since it has been regularized
and is finite, we may take it as a new input and continue the
iteration. However, it does not converge, The iteration results
oscillate between $\sum_{pv}\sim k^{+3}$ and $\sum_{pv}\sim k^{+1}$.
This is owing to the fact that different regularization will render
the integral equation different. Following the above naive iteration
procedure, one is actually solving two integral equations: one with
$G(k)\sim k^{-3}$ and the other with $G(k)\sim k^{-1}$ in
Eq.(\ref{eq4}). Therefore, one has to look for a formulation which
requires only one regularization for the entire iteration process.
This can be achieved by the K$\ddot{a}$llen-Lehmann spectral
representation, which for $G(k)$ can be written as
\begin{subequations}
\renewcommand{\theequation}{\theparentequation-\arabic{equation}}
\begin{align}
 G(k)&=-Z_t \int_0^{\infty } d m^2\frac{\gamma_\mu k_\mu f_{\alpha}(-m^2)+i M_t
f_\beta(-m^2)}{k^2+m^2-i\varepsilon},\label{eq9.1}\\
f_\gamma(-m^2)&=\delta(m^2-M_t^2)+\theta(m^2-m_1^2)\gamma(-m^2),\label{eq9.2}
\end{align}
\end{subequations}
where $(-Z_t)$ is the residue of G(k) at the physical pole
$\gamma_\mu k_\mu=iM_t$, $\gamma=\alpha\  or \ \beta$,
$m_1=M_t+m_\pi$ and $\theta$ is the step function. Substituting
Eq.(9) into Eq.(\ref{eq4}), choosing $\Delta_\pi(q)=\Delta^0_\pi(q)$
and again using Feynman's parametrization and dimensional
regularization, we find
\begin{subequations}
\renewcommand{\theequation}{\theparentequation-\arabic{equation}}
\begin{align}
a_R(k^2)=&\frac{3f^2Z_t}{16\pi^2}\int_0^{\infty}dm^2f_\alpha(-m^2)\int_0^1dx
\left\{\left[(1+3x)\ln K^2-2x\right]K^2+x^2(1-x)k^2\ln
 K^2\right\},\label{eq10.1}\\
b_R(k^2)=&\frac{3f^2Z_tM_t}{16\pi^2M}\int_0^{\infty}dm^2f_\beta(-m^2)\int_0^1dx
\left[x^2k^2\ln K^2+K^2(1-2\ln K^2)\right],\label{eq10.2}\\
K^2=& x(1-x)k^2+x m^2+(1-x)m_\pi^2, \label{eq10.3}
\end{align}
\end{subequations}
where to regularize $a $ and $b$ we have used $\overline{MS}$
(modified minimal subtraction) and deleted terms proportional to
$(1/\varepsilon-\gamma_E+\ln4\pi)$. From Eqs.(\ref{eq7}) and (9))
one can further derive the following relations
\begin{align}
Z_t\alpha(k^2)&=\frac{1}{\pi}Im\frac{1+a_R(k^2)}{D(k^2)},\quad
Z_t\beta(k^2)=\frac{1}{\pi}\frac{M}{M_t}Im\frac{1+b_R(k^2)}{D(k^2)},\nonumber\\
D(k^2)&=k^2\left[1+a_R(k^2)\right]^2+M^2\left[1+b_R(k^2)\right]^2.
\label{eq11}
\end{align}
We still need to know how to determine M and $Z_t$. By definition
and Eq.(\ref{eq7}) one easily finds
\begin{equation}
\begin{split}
&M_t\left[1+a(-M_t^2)\right]-M\left[1+b(-Mt^2)\right]=0,\\
&Z_t^{-1}=1+a(-M_t^2)+2M_t\left[Mb'(-Mt^2)
 -M_ta'(-M_t^2)\right],\label{eq12}
\end{split}
\end{equation}
where $C'(k^2)\equiv d C(k^2)/dk^2$. Eqs.(10)-(\ref{eq12}) build a
closed set of equations for the determination of $a_R, b_R$, $\alpha
$ and $\beta$. They can be solved by the method of iteration.
However, we have found that the iteration procedure does not
converge, though $\overline{MS}$ has been proved successful in other
aspects. In the following we shall describe our solution to
Eq.(\ref{eq4}) obtained by the counterterm method. The counterterm
CTL to be included in the Lagrangian can be written as
\begin{equation} CTL=-\overline{\psi}\left[M_c
+\Sigma_{l=1}^3\frac{1}{l!}\eta_l (\gamma_\mu
\partial_\mu)^l\right]\psi. \label{eq13}
\end{equation}
Its contribution to the fermion self-energy CTS is given by
\begin{subequations}
\renewcommand{\theequation}{\theparentequation-\arabic{equation}}
\begin{align}
CTS&=i\left[M_c +\Sigma_{l=1}^3\frac{1}{l!}\eta_l (i \gamma_\mu
k_\mu)^l\right] ,\label{eq14.1}\\
\Sigma_R(k)&=\Sigma(k)+CTS =\gamma_\mu k_\mu a(k^2)-i M b(k^2)+CTS,
\label{eq14.2}
\end{align}
\end{subequations}
where $\Sigma_R(k)$ means the renormalized self-energy. We note that
for our purpose there is no need to redefine $(\gamma_\mu
\partial_\mu)^l$ in Eq.(\ref{eq13}), because the contribution of CTL can always be made to form
 a pair with the self-energy $\Sigma(k)$ and
so its net effect is included in $\Sigma_R(k)$ (see FIG.\ref{fig1}).
Hereafter the counterterm method will be referred to as the method
of renormalization. In order to determine the parameters in
Eq.(\ref{eq13}), we shall use the on-shell renormalization
conditions which read
\begin{subequations}
\renewcommand{\theequation}{\theparentequation-\arabic{equation}}
\begin{align}
\Sigma_R(k)|_{\not k=i M_t}&=\frac{\partial\Sigma_R}{\partial\not k}\Big |_{\not k=i M_t}=0, \label{eq15.1}\\
\frac{\partial^2\Sigma_R}{\partial \not k^2}\Big |_{\not k=i M_t}&=i
\kappa; \quad \frac{\partial^3\Sigma_R}{\partial \not k^3}\Big
|_{\not k=i M_t}=\lambda, \label{eq15.2}
\end{align}
\end{subequations}
where ${\not k}\equiv\gamma_\mu k_\mu$. The two conditions in
Eq.(\ref{eq15.1}) imply $M=M_t$  and $ Z_t=1$. Usually one also put
$\kappa=\lambda=0$. Here we leave them to be two constant free
parameters which may be determined by other requirements or by
fitting experimental data. we have the following relations
\begin{equation}
\begin{split}
a_R(k^2)&=\hat{a}_R(k^2)+M_t\kappa-\frac{1}{2}\lambda
M_t^2+\frac{1}{6}\lambda k^2\\
b_R(k^2)&=\hat{b}_R(k^2)+\frac{1}{2} M_t\kappa-\frac{1}{6}\lambda
M_t^2-\frac{1}{2}\frac{\kappa-\lambda M_t}{M_t} k^2, \label{eq16}
\end{split}
\end{equation}
where $\hat{a}_R$ and $\hat{b}_R$ are the results for
$\kappa=\lambda=0$. Their explicit expressions are
\begin{subequations}
\renewcommand{\theequation}{\theparentequation-\arabic{equation}}
\begin{align}
\hat{a}_R(k^2)&=-\frac{3f^2}{16\pi^2}\int_0^{\infty}dm^2
f_\alpha(-m^2) \int_0^1dx
\left\{\left[x^2(1-x)k^2+(1+3x)K^2(k^2)\right]\ln\frac{K^2(-M_t^2)}{K^2(k^2)}\right\}\label{eq17.1}\\
&+(k^2+M_t^2)C_\alpha+4M_t^2(2M_t^2-k^2)C_\alpha(1)-2M_t^2
(M_t^2-k^2)C_\alpha(2)+4M_t^4(M_t^2-1/3k^2)C_\alpha (3),\nonumber
\end{align}
\begin{align}
\hat{b}_R(k^2)&=-\frac{3f^2}{16\pi^2}\int_0^{\infty}dm^2f_\beta(-m^2)
\int_0^1dx
\left\{\left[2K^2(k^2)-x^2k^2\right]\ln\frac{K^2(k^2)}{K^2(-M_t^2)}\right\}-(k^2+M_t^2)C_\beta
\nonumber\\
&-M_t^2(10k^2-2M_t^2)C_\beta(1)
+4M_t^2k^2C_\beta(2)-4M_t^4(k^2-1/3M_t^2)C_\beta(3),\label{eq17.2}
\end{align}
\end{subequations}
\begin{figure}[b]
\scalebox{0.6}[0.7]{\includegraphics{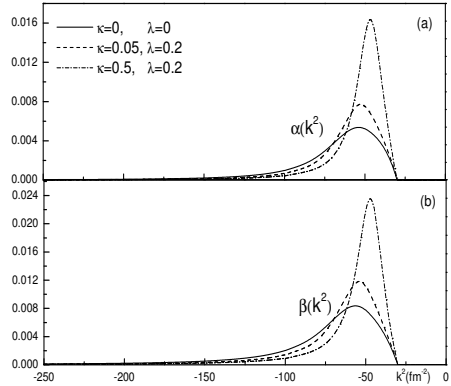}}
\caption{\label{fig2}Baryon spectral functions for $\pi$ model with
pseudovector $\pi-N$ coupling: (a) $\alpha(k^2)$ and (b)
$\beta(k^2)$.}
\end{figure}
\begin{figure}
\scalebox{0.6}[0.7]{\includegraphics{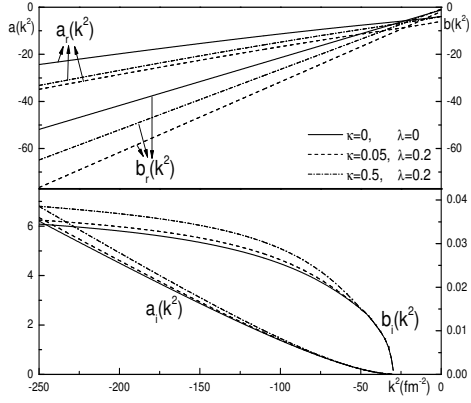}}
\caption{\label{fig3}Self-consistent results of $a(k^2)$ and
$b(k^2)$ for $\pi$ model with pseudovector $\pi-N$ coupling. top:
the real part; bottom: the imaginary part. $a(k^2)$ and $b(k^2)$
refer to left scale and  right scale, respectively.}
\end{figure}
where $C_\gamma$ and $C_\gamma(l)$ ($\gamma=\alpha$ or $\beta$,
$l$=1 to 3) are $k^2$-independent constants. Their explicit
expressions can be found easily from Eqs.(14) and (15). As they are
long and space consuming, we shall not write them down here.
Eqs.(11),(\ref{eq16}) and(17) now
 build a closed set of equations. We have solved this set by
 iteration. The initial input ia taken as $\alpha=\beta=0$, It is
 found that the iteration series converges quite quickly. In
 FIG.\ref{fig2} and FIG.\ref{fig3} we have plotted our numerical results for the
 self-consistent sets ($\alpha,\beta$) and ($a_R,b_R$). Besides
 $\kappa=\lambda=0$, we have further calculated two cases of
 $\kappa\neq0$ and $\lambda\neq0$ as an illustration. From FIG.\ref{fig2} one
 observes that through appropriate choice of their values the peak
 can be made sharper and more pronounced. Thus, the introduction of
 these two additional parameters is physical meaningful and worth
 considering. Finally we would like to emphasize that the
 solution to the regularized integral equation, Eq.(\ref{eq4}), exists and
 can be obtained by the above iteration procedure also offers a
 noteworthy support of our proposal for the extension of
 renormalizability.

We would like to point out that the above results of
 renomalizability for PVI are general, namely if an interaction is
 nonrenormalizable in OS, it is always possible to find a
 DS such that it becomes renormalizable. It is interesting to note
 that the possibility comes from the condition $r_{nr}<0$. Let
 $i=\lambda$ be a NRI and use a superscript * to label OS. Consider
 a lowest order fermion self-energy diagram $\Sigma_F(k)$ with each
 of its external vertices A and B (see FIG.\ref{fig1}) being a $\lambda$.
 According to Eq.(3) the asymptotic behavior of $\Sigma_F(k)$ can be
 written as
$\tilde{d}_F^*=d_F^*+{\Sigma}'d_i=2\left[p_f^*+|r_\lambda^*|\right],$
because $r_\lambda^*<0$ and $E_f=N_\lambda=2$. Since regularization
will not affect $\tilde{d}_F^*$ (see, for instance, Eqs(10) and
Eqs.(\ref{eq16},17)), if we substitute the regularized $\sum_F(k)$
for $\sum_d(k)$ in Eq.(\ref{eq2.2}), its solution $G_{d F}(k)$, as
shown by Eq.(\ref{eq7}), behaves asymptotically as
$k^{-\tilde{d}_F^*}=k^{-2p_{_{d f}}}$  or $p_{_{d
f}}=\tilde{d}_F^*/2$. From Eq.(\ref{eq3.3}) one easily finds that in
DS $r_\lambda=r_\lambda^*+n_{_{\lambda
f}}\left(\tilde{d}_F^*/2-p_f^*\right)=(n_{_{\lambda
f}}-1)|r_\lambda^*|$, which shows $r_\lambda\geq0$, because
$n_{_{\lambda f}}\geq1$. Thus, NRI $\lambda$ becomes renormalizable
or super-renormalizable in DS. If $\lambda$ contains no fermion
fields, clearly we may instead consider a lowest order boson
self-energy diagram $\Pi_B(k)$ with each of its external vertices A
and B being $\lambda$. Following the same argument, we again find
that the asymptotic behavior of $\Pi_{_B}(k)$ is given by
$\tilde{d_B^*}=2(p_{_B}^*+|r_\lambda^*|)$, and the dressed
propagator $\Delta_{dB}(k)$ behaves asymptotically as
$k^{-\tilde{d}_{_B}^*}=k^{-2p_{_{d B}}}$  or  $p_{_{d
B}}=\tilde{d}_{_B}^*/2$, i.e. in DS determined by $\Delta_{_{d
B}}(k)$ $r_\lambda=r_\lambda^*+n_{_{\lambda
B}}\left(\tilde{d}_{_B}^*/2-p_{_B}^*\right)=(n_{_{\lambda
B}}-1)|r_\lambda ^*|.$ Thus $r_\lambda\geq0$, which again confirms
our above conclusion. Clearly the above argument also applies to the
case of more than one NRI. Say we have $\xi$ different interactions
which can contribute to the fermion self-energy and among which
there are two NRIs $\lambda$ and $\eta$ with
$r_{\lambda}^*<r_{\eta}^*$ or $|r_{\lambda}^*|>|r_{\eta}^*|$.
Altogether we can build $\xi
$$\times $$\xi$ fermion self-energy insertions $\Sigma_{ij}(k)$ with
the two external vertices being interaction $i$ and $j$,
respectively. $\Sigma_{ij}(k)$ should be regularized, if it is
divergent. Now set $\Gamma_d(k)=\Sigma_{i,j=1}^\zeta\Sigma_{ij}(k)$
in Eq.(2), where $\zeta$ may be smaller than $\xi$. The choice of
$\zeta$ will affect the efficiency of calculation, but is irrelevant
to our present discussion. We shall only require that
$i$=$j$=$\lambda$ is included in $\Gamma_d(k)$, From Eqs.(3) and
(\ref{eq7}) one finds that the asymptotic behavior of $G_{dF}(k)$ is
given by $k^{-d_F^*}=k^{-2p_{_{df}}}$ with
$p_{_{df}}=p_f^*+|r_{\lambda}^*|$. Thus, in DS
$r_{\lambda}=(n_{_{\lambda f}}-1)|r_{\lambda}^*|\geq0$, while
$r_{\eta}=n_{_{\eta f}}|r_{\lambda}^*|-|r_{\eta}^*|>0$. Clearly our
conclusion holds generally. Moreover one observes that all the
renormalizable interactions in OS become super renormalizable in the
above DS, if the latter is constructed by means of NRI as shown
above. Note that DS is derived from OS in a simple and natural way.
As we have emphasized, the reason that they may differ significantly
in their property of renormalizability is because of regularization
and nonrenormalizability. We have demonstrated that DS  can be
constructed by dressed propagators determined by the Dyson-Schwinger
equation with the regularized fermion or boson self-energy as its
kernel. It is seen that besides being renormalizable, DS further
offers a non-perturbative method for the calculation. To show that
DS exists, we have only made use of Eq.(3) and the existence of
regularized expression for the self-energy. Thus, we may conclude
that the present quantum field theory based on special relativity
with interactions not too exotic is actually a renormalizable
theory, if a proper framework of representation is established. A
more stringent condition is that the solution to the self-energy
integral equation should exist. In principle this would not be a
problem, if it were not for the fact that regularization is
necessary.Though a general mathematical existence proof is beyond
the scope of this letter, in text we have suggested a method which
can be used to study and check each special case individually.

This work is supported by the National Nature Science Foundation of
China with grant number 10375026.

\end{document}